\def\erg{{\rm\thinspace erg}}

\def\keV{{\rm\thinspace keV}}

\def\Msun{\hbox{$\rm\thinspace M_{\odot}$}}
\def\pc{{\rm\thinspace pc}}

\def\s{{\rm\thinspace s}}
\def\yr{{\rm\thinspace yr}}

\def\ergps{\hbox{$\erg\s^{-1}\,$}}

\input{aipcheck}

\documentclass[
    ,final            
  ]
  {aipproc}

\layoutstyle{6x9}

\begin{document}

\title{Heating and Cooling in Clusters and Groups}

\classification{\texttt{ 98.65Cw, 98.65Hb}}

\author{A.C. Fabian}{
  address={Institute of Astronomy, Madingley Road, Cambridge CB3 0HA, UK}
}

\author{J.S. Sanders}{
  address={Institute of Astronomy, Madingley Road, Cambridge CB3 0HA, UK}
}

\begin{abstract}
  The gas in the cores of many clusters and groups of galaxies has a
  short radiative cooling time. Energy from the central black hole is
  observed to flow into this gas by means of jets, bubbles and sound
  waves. Cooling is thus offset by heating. We discuss the mechanisms
  involved and observed in the X-ray brightest clusters and explore
  the closeness of the heating/cooling balance. It is surprisingly
  tight on the cooling side when soft X-ray spectra are
  examined. Non-radiative cooling by mixing is suggested as a means to
  relax the apparent strong lack of cooling. Nevertheless the heating
  and cooling must balance on average to better than 20 per cent.
\end{abstract}

\maketitle


\section{Introduction}
The hot gas in the cores of more than one third of clusters of
galaxies, and most elliptical-rich groups of galaxies, has a radiative
cooling time shorter than about 3~Gyr. This is significantly less than
the age of these systems, so the gas would develop into a cooling flow
in the absence of additional heating \citep{Fabian94}.  X-ray
observations show that the central gas temperature drops and the
density consequently rises (Fig.~1), as expected in a cooling flow,
but the temperature does not drop completely. (They are therefore
known as cool-core clusters.) The likely explanation is that the
active nucleus of the central galaxy injects energy into its
surroundings which offsets cooling. The activity is evident in jets
seen in radio observations and bubbles inflated by the jets seen in
X-ray images (for reviews see \cite{PetersonFabian06, McNamaraNulsen07}).

A study of a small complete sample of clusters by \cite{DunnFabian06}
shows that 90 per cent of those with a central cooling time less than
3~Gyr have a central radio source and 70 per cent have a bubbles
associated with that source (Fig.~2). In other words, where cooling is
proceeding most rapidly then there is a high probability that energy
is being injected. Moreover, the duty cycle of this activity is high,
between 70 and 90 per cent.

The power required by the heat source ranges to above $10^{45}\ergps$
in the most extreme cases, but is more typically $10^{44}\ergps$ for a
cluster and $10^{43}\ergps$ for a group.  The energy in a bubble of
relativistic gas with volume $V$ is $\sim4PV$, where $P$ is the
surrounding pressure, and the buoyancy time, which is the time taken
to fill the bubble, is comparable to the sound crossing time of the
bubble. The power can then be obtained from the X-ray data, yielding
estimates which in most cases are comparable to the cooling rate of
the surrounding cool core \cite{Birzan04, Dunn05, Rafferty06}.  The bubbles
demonstrate that jets transport considerable energy in a radiatively
inefficient manner, with less than 0.1 to 0.01 per cent of the power
emerging as radiation.

The right amount of energy is being injected, so energy is not a
problem, but how does it get distributed? How close is the
heating/cooling balance? Observations suggest that it is better than
10 per cent in some objects. How does such tight feedback work?
Particularly when the wide range of timescales involved from several
Gyr down to less than $10^8\yr$ for the coolest gas seen below 1~keV
is considered. We see no gas at all in the X-ray band at temperatures
below 5 million K, how is the gas immediately hotter than this
prevented from cooling?  To examine these issues we use the best X-ray
data from the brightest clusters.


\begin{figure}[h]
  \includegraphics[height=.3\textheight]{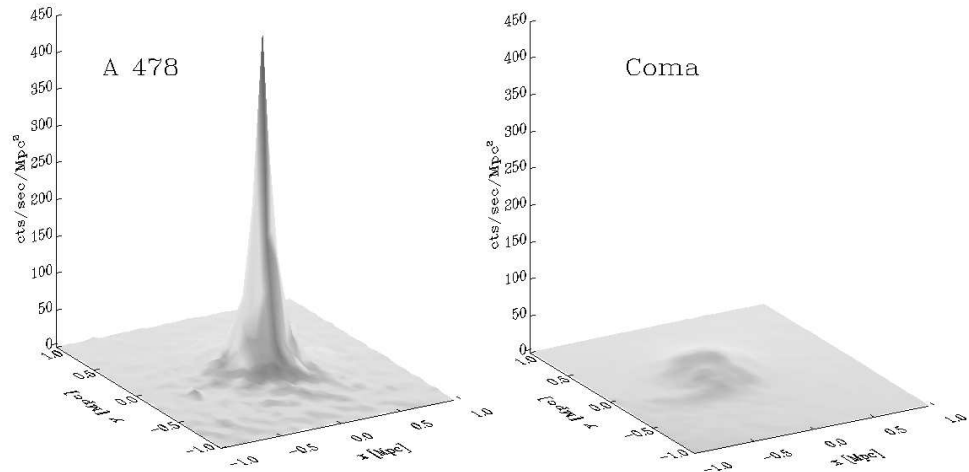}
  \caption{X-ray surface brightness of a cool-core cluster A478 (left)
    and a non cool-core cluster, the Coma cluster, on the right
    (courtesy of Steve Allen). They have been adjusted as if at the
    same distance. The emission depends on the density squared so is
    strongly peaked where the density rises and the temperature
    drops.}
\end{figure}
\begin{figure}
  \includegraphics[height=.3\textheight]{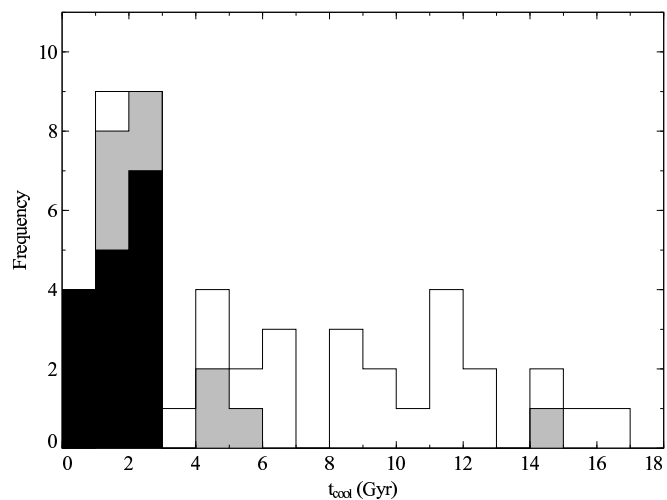}
  \caption{Histogram of central cooling times of clusters in the
    Brightest 55 sample \citep{Dunn05}. Black indicates the
    presence of bubbles, grey indicates a radio source. }
\end{figure}

\section{Distributing the energy}

\begin{figure}
  \includegraphics[height=.3\textheight]{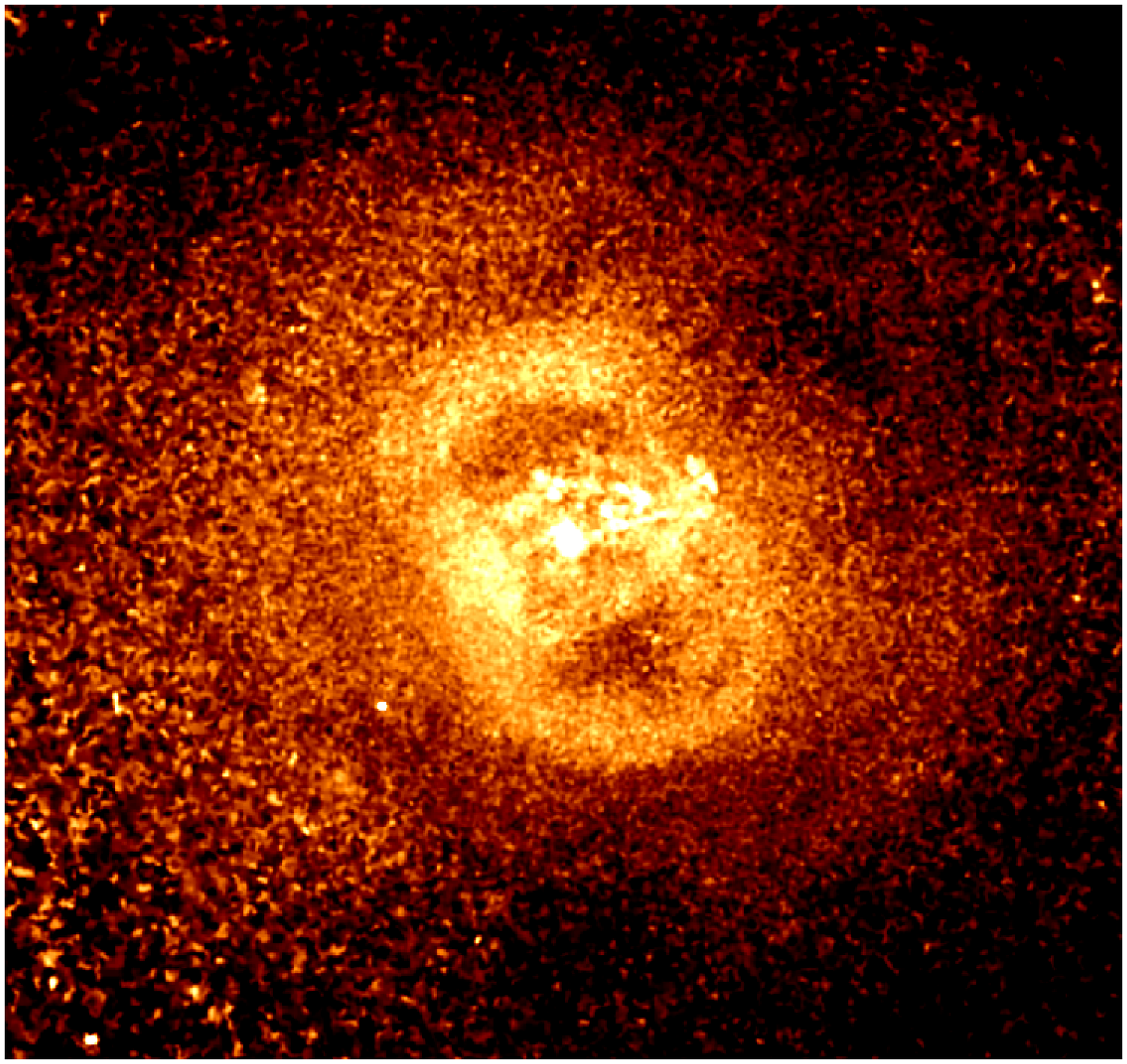}
 \includegraphics[height=.3\textheight]{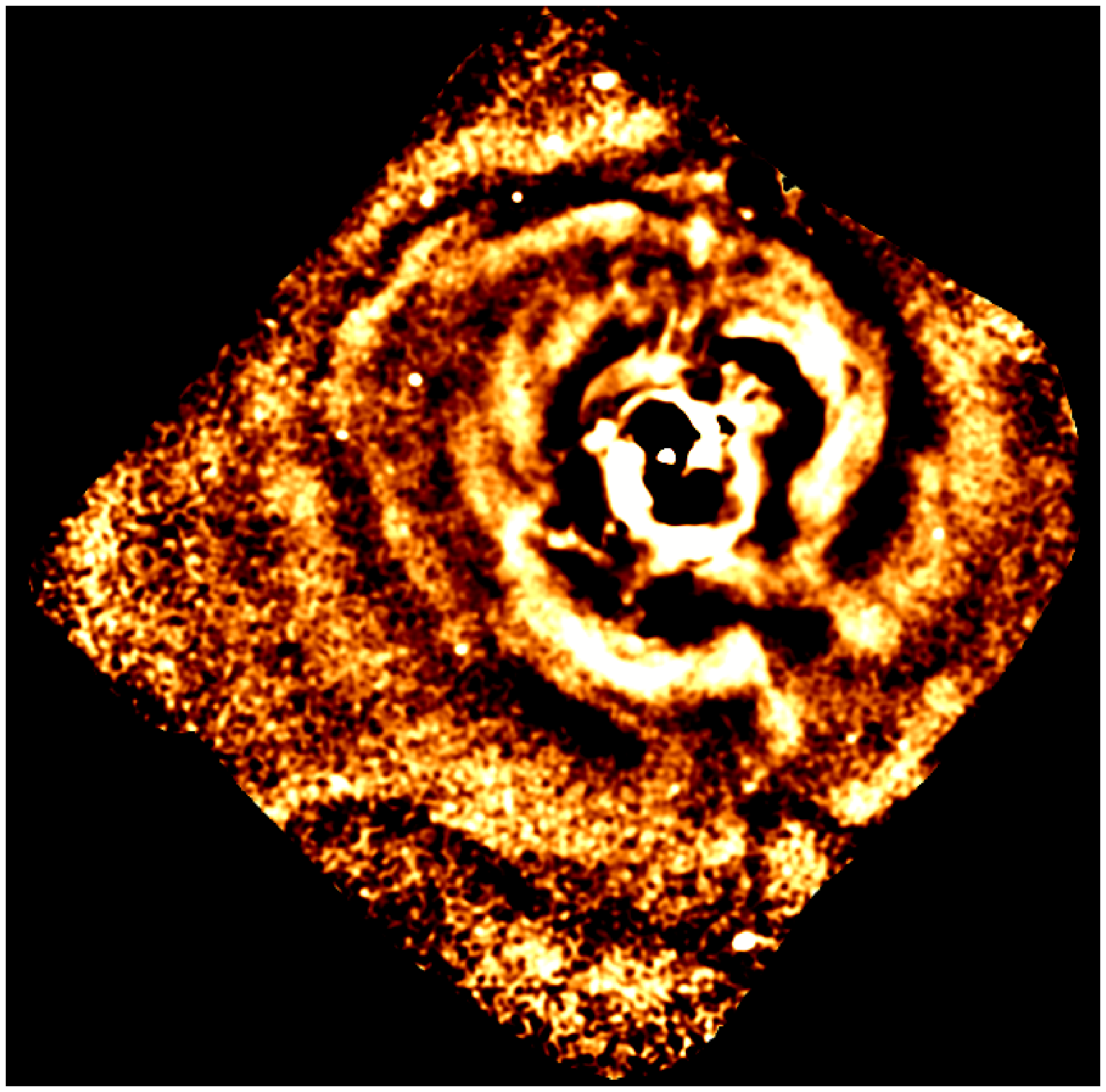}
  \caption{Left: Pressure map showing the central bubbles around NGC\,1275
    in the Perseus cluster. Right: unsharp masked image showing
    ripples. The bubbles seen in the left image correspond to the two
    dark regions either side of the nucleus.  }
\end{figure}

The X-ray brightest cool-core cluster, the Perseus cluster, also has
the deepest Chandra exposure of nearly 1Ms, with about 70 million
photons contributing to the image. It shows clear high pressure shells
around the two radio bubbles (Fig.~3) and at least two buoyant outer
bubbles \citep{FabianPer06}. There are weak shocks (Mach number 1.3)
at the edge of the shells which presumably propagate outward
developing into sound waves. Such sound waves explain the sequence of
ripples seen in the X-ray surface brightness image
(\cite{FabianPer03}, \cite{SandersPer07}, Fig.~3). There should be at
least one sound wave per bubble inflation/buoyant rise cycle; large
fluctuations in the jet power could lead to more.

The energy flux in the sound waves is seen to be close to that
required to counterbalance the radiative cooling. Thus provided the
sound energy is dissipated in the gas, we are seeing a mechanism which
converts the directional jet energy into a quasi-isotropic heat
source. Dissipation depends on viscosity, which is difficult to
estimate in the intracluster medium. This is because the gas is
magnetized, probably in a tangled manner. A bulk viscosity may result
if the application of pressure changes the magnetic
configuration. Otherwise the maximum shear viscosity is the
Spitzer-Braginsky value (which is high) with the true value reduced by
the magnetic fields.

Several observational clues argue that the viscosity may be high.  The
quasi-linear nature of the H$\alpha$ optical emission filaments seen
around the central galaxy of the Perseus cluster, NGC\,1275, and the
Brightest Cluster Galaxies (BCGs) of many other cool-core clusters
(Fig.~4) suggest that the gas is not highly turbulent. The SW X-ray
plume seen near M87 has a very sharp edge, again showing that flows in
the gas must be quasi-laminar and not chaotic. Estimates of the
outward diffusion rates of the iron-rich gas around BCGs \citep{RebuscoDiff05,
Graham06} also argue that motions are relatively quiet, and thus that
the gas may be viscous.

\begin{figure}
  \includegraphics[height=.3\textheight]{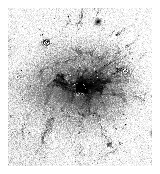}
  \includegraphics[height=.3\textheight]{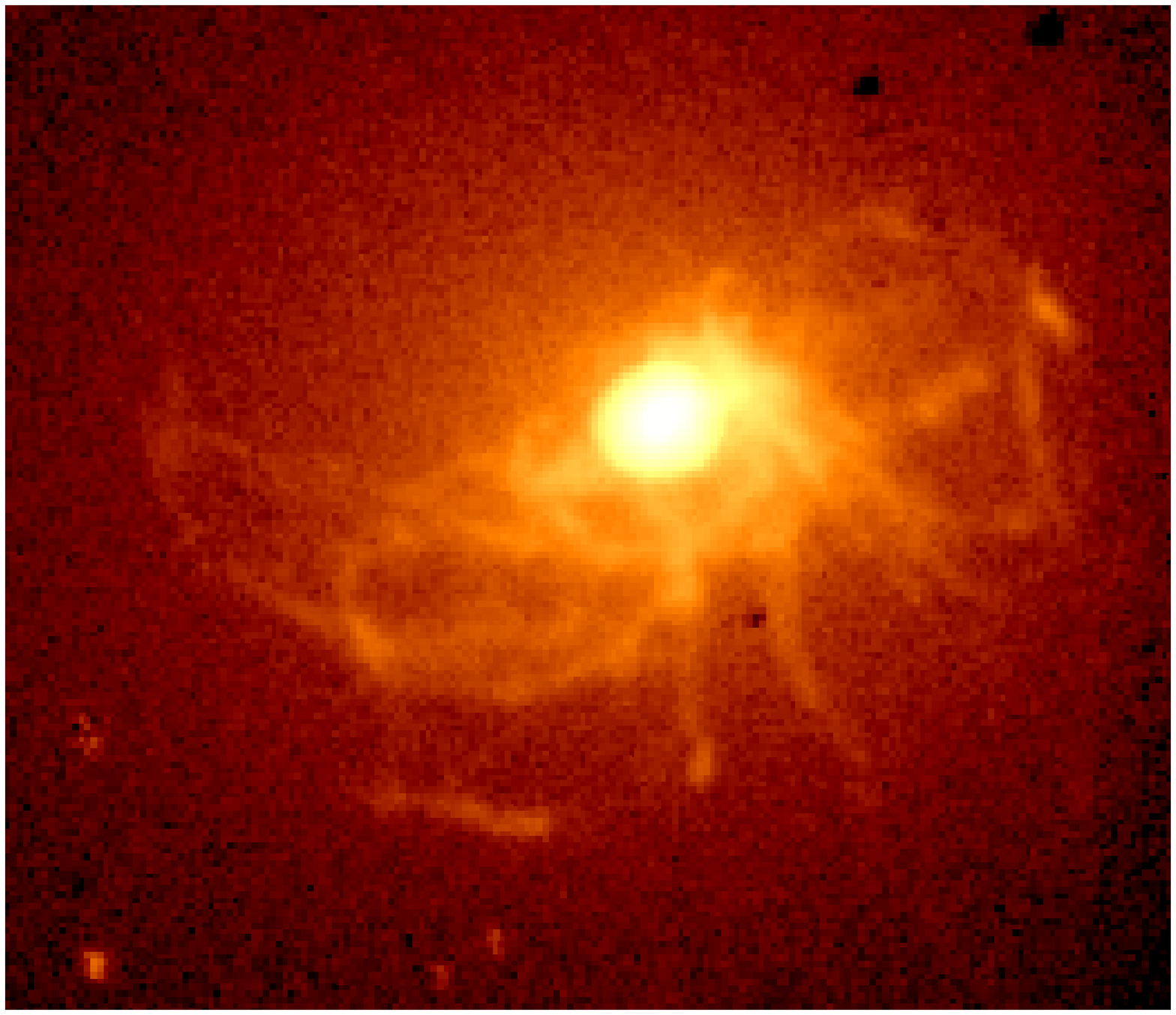}
  \caption{Left: HST image of the H$\alpha$ filaments around NGC\,1275
    \cite{FabianFilament08}. The
    image is about 50~kpc from top to bottom. Right: H$\alpha$ image
    of NGC\,4696 in the Centaurus cluster \cite{Crawford05}.}
\end{figure}

Although the sound waves seen in the Perseus cluster diminish with
radius out to 100~kpc, consistent with energy being dissipated, outer
ripples appear stronger. This could be due to the central source
having been a few times more energetic a few 100 Myr ago. This would
also explain the very long (50~kpc) Northern H$\alpha$ filaments, if
they were dragged out by rising bubbles at that time.

Ripples and weak shocks are also found in the Virgo
\citep{FormanM8707}, Centaurus \citep{SandersCen08} and A2052
(Blanton, this meeting) clusters, all of which are X-ray bright and
have deep Chandra data. They will be difficult to see in general
without very good data \citep{Graham08}. Nevertheless they appear to
be a major channel by which energy from the black hole is distributed
in clusters and groups.

\section{Cooler gas}

We can address the issue of the tightness of feedback by looking for
evidence of overheating or underheating. If too much energy is
injected then we might expect evidence for strong turbulence or
heating. Some non-cool core clusters may have resulted from this (e.g.
\cite{Donahue05, Osullivanawm705}).  We can say that it is not common
but it is unclear what we would look for. If jets become more powerful
then they can turn into FR\,II sources with energetic hotspots at the
end of rapidly-extending long jets, such as seen in Cyg\,A and
3C295. Both of these objects are in cool core clusters. Their hotspots
are expected to rapidly pass beyond the inner region where heat is
needed. They may then be relatively inefficient in heating most of the
core gas (i.e. that not along the jet direction). MS\,0735.6 is a
good example where a very energetic event, or sequence of events, has
dumped most of the energy well beyond the core \citep{McNamara09}.

The luminous quasar H1821+643 lies at the centre of a rich
cluster. Chandra imaging shows little impact of such powerful fuelling
of the black hole on the gas in the core beyond 20~kpc (Russell et al,
submitted).

If too little energy is supplied then radiative cooling ensues and we
can expect to accumulate cold gas and possibly have star formation.
Considerable cold gas is associated with many cool core cluster BCGs,
mostly in the form of molecular gas \citep{Edge01, Salome06}. The
filaments of Perseus (Fig.~4) are mostly molecular, even the outer
ones \citep{HatchPer06, Salome06}, with a total mass of about
$10^{11}\Msun$. A plausible scenario for the origin of the filaments
is that they are due to cooled gas, enriched by stellar mass loss
(including dust), which is then dragged out beneath buoyantly-rising
bubbles.

The filaments are very thin and likely magnetized
\citep{FabianFilament08}. They are quite luminous and have a
lowly-ionized spectrum which can be explained by heating by particles,
either low-energy cosmic rays within the filaments themselves perhaps
energised by their motion, or by the outside hot gas \citep{Ferland09}.

Star formation is common in such BCGs (\cite{ODea08} and refs therein)
with, for example, A1835 showing the highest star formation rate for
an early-type system in the low redshift universe. The star formation
rate is however only about 10 per cent of what could result if a
steady-state cooling flow took place without any heating.

We have been using spectra from the XMM Reflection Grating
Spectrometer to quantify the amount of gas cooling below 10 million K
in a number of groups and clusters. For the Centaurus cluster we see
clear evidence of gas down to about 4~MK but not below that
temperature (Fig.~5; \cite{SandersRGS08}), with similar results for
some groups (Fig.~6, Sanders 2009 submitted). A similar conclusion has
been obtained on M87 by \cite{Werner06}. No group or cluster shows
clear evidence for gas cooling below that temperature. In particular,
no OVII emission is seen (as expected from gas at say 3~MK), although
the constraint does not rely just on that species.

The limit on cooling in the Centaurus cluster is so stringent (Fig.~5,
right) that one could conclude that the heating/cooling balance is
kept to within a few per cent. (The abundance gradient mentioned
earlier restricts any overheating.)

\begin{figure}
  \includegraphics[height=.3\textheight]{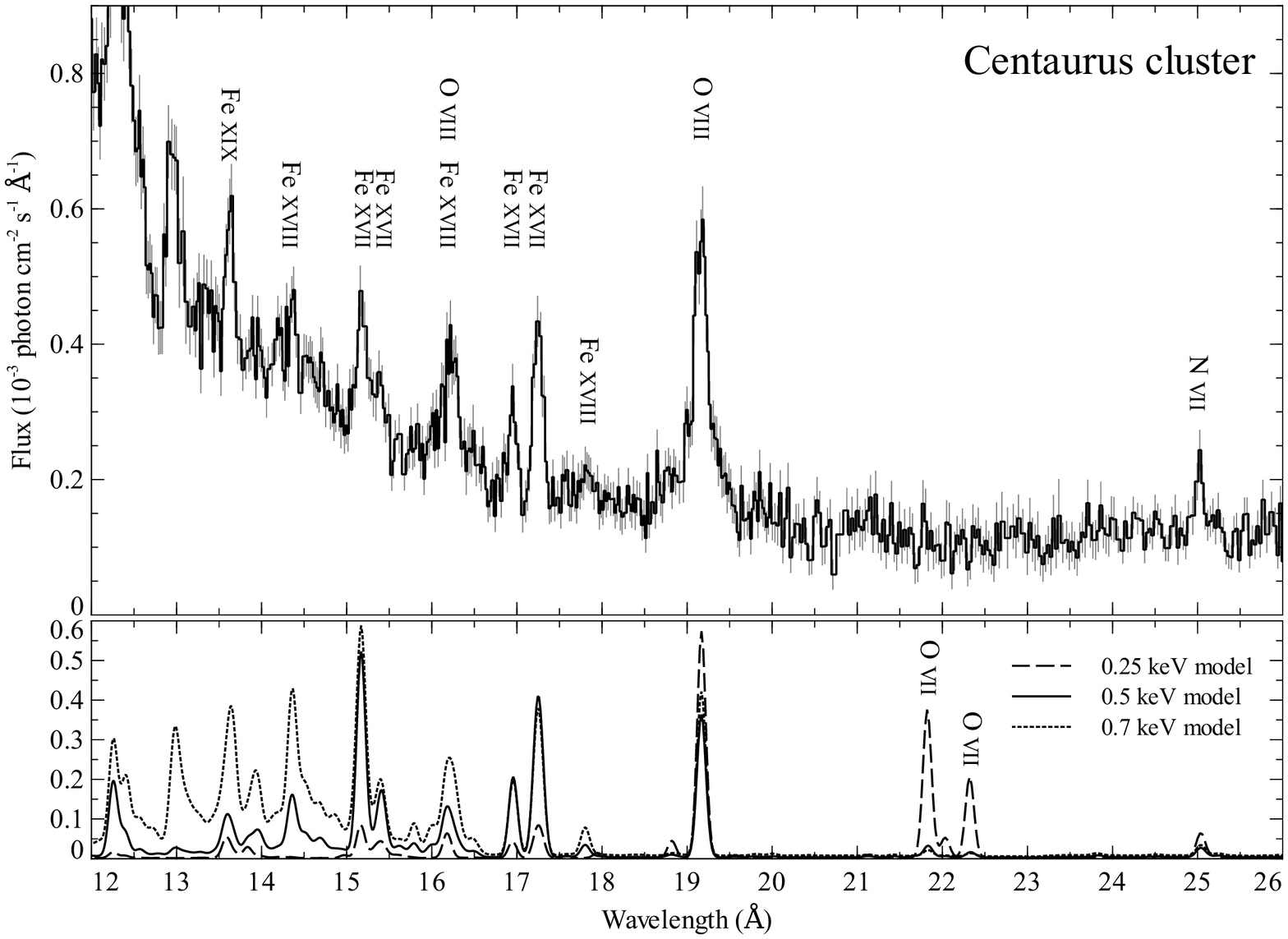}
  \includegraphics[height=.3\textheight]{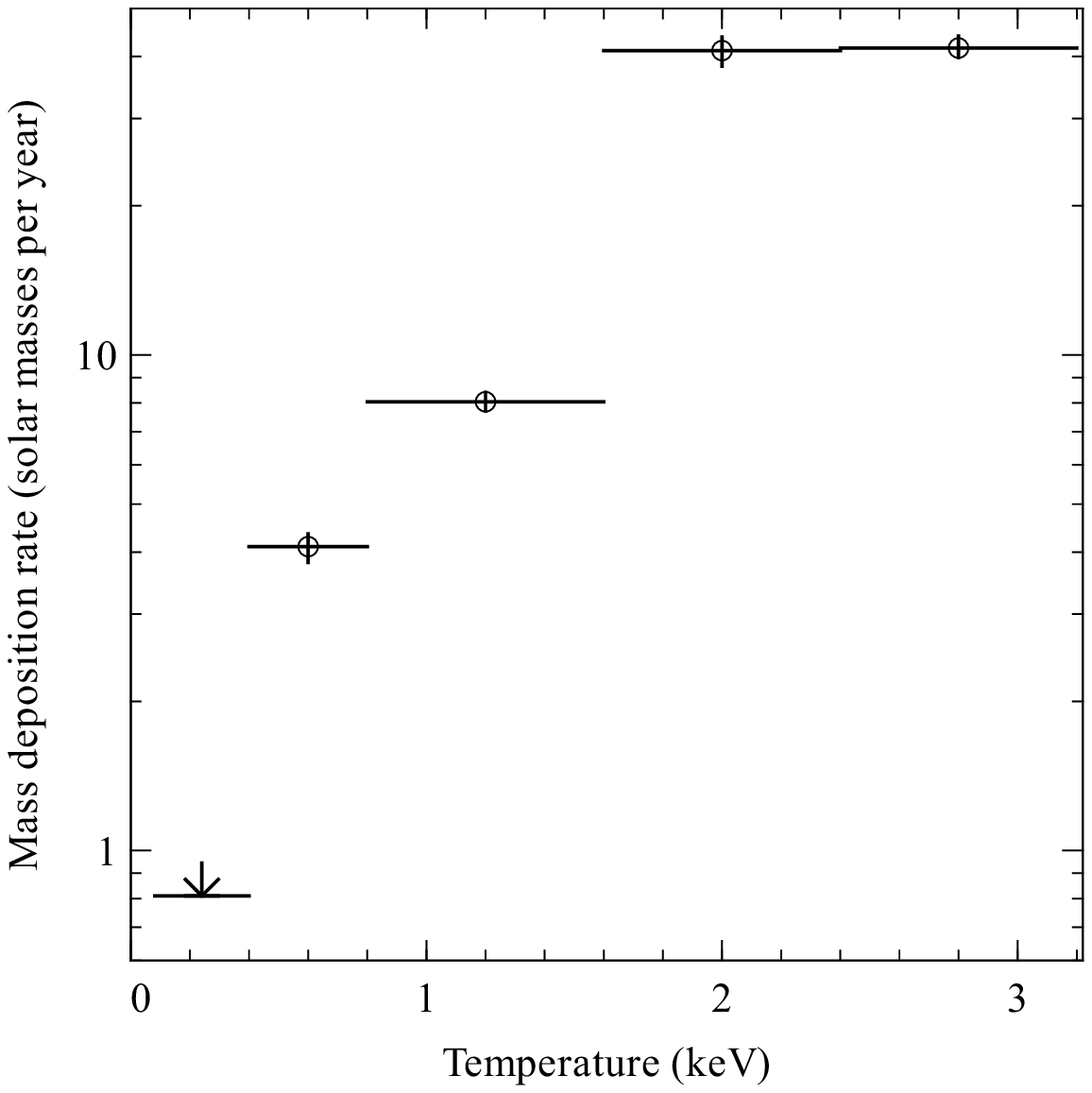}
  \caption{Left: XMM RGS spectrum of the centre of the Centaurus
    cluster \citep{SandersRGS08}. Comparison with the lower panel
    shows that there is little gas seen below 0.5~keV. Right: Mass
    cooling rate diagram emphasising the absence of lower temperature gas.}
\end{figure}
\begin{figure}
  \includegraphics[height=.3\textheight]{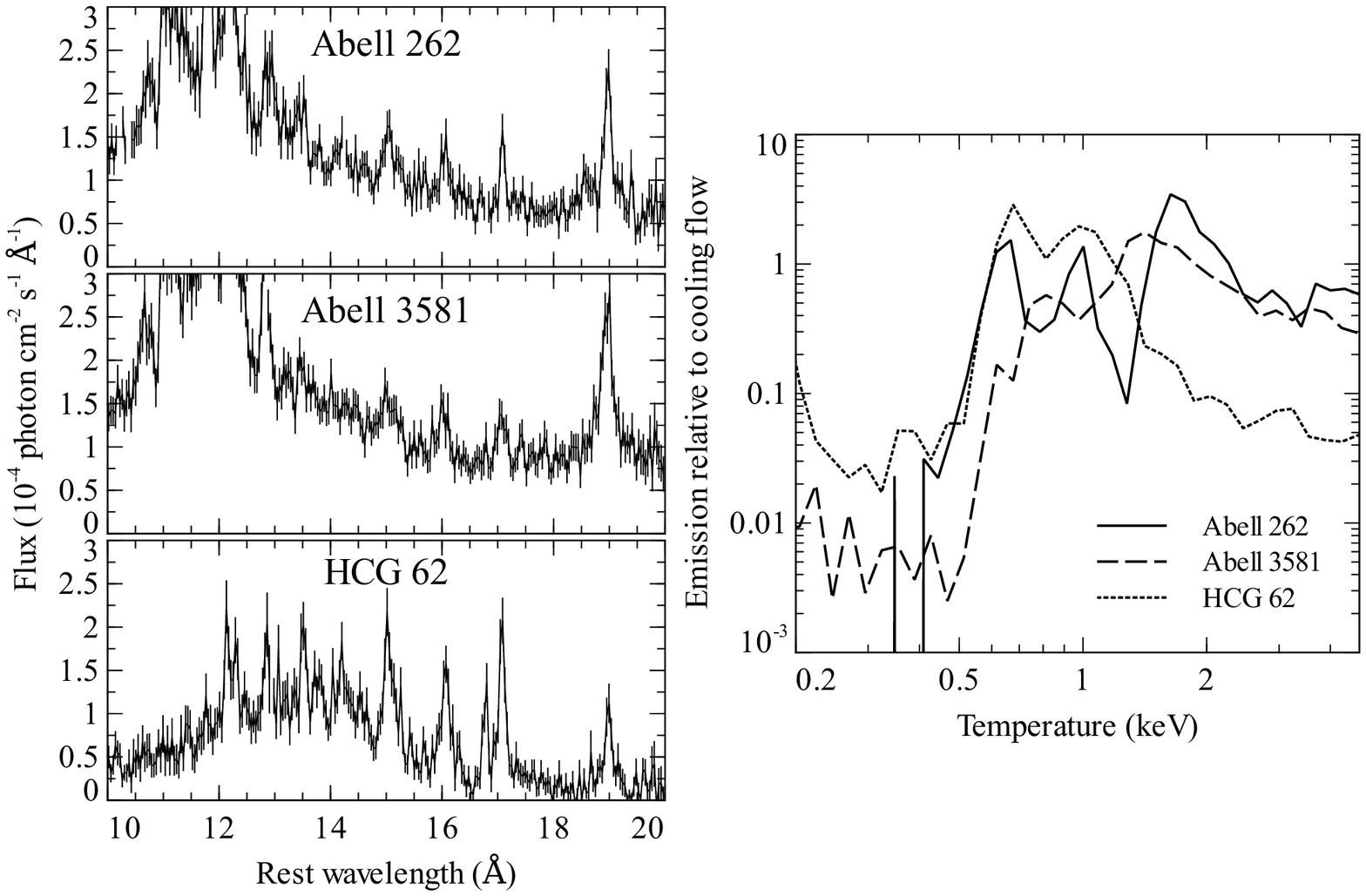}
  \caption{XMM RGS spectra of 3 groups (Sanders et al 2009, submitted)
    with relative emission measures shown to the right. Again there is
    a lack of cooler gas.}
\end{figure}

The coolest gas ($<1\keV$) in the Centaurus cluster seen in the
Chandra image is clumpy. It is also the most rapidly cooling gas with
a cooling time of only $10^7\yr$. If heating explains the lack of yet
cooler X-ray emitting gas, then it must be very selectively targeted
by the heat source. An alternative is that the coolest gas does cool,
but non-radiatively. It cools by mixing with the cold filamentary gas
which is distributed over the same inner, few kpc, region. Most of the
thermal energy of the hotter gas is then eventually radiated in the
infrared band by dust and molecular emission (\cite{FabianCFlow02},
\cite{Soker04}).

This relaxes the balance to be within 10 to 20 per cent, with the
constraints on cold gas and star formation (which may be sporadic)
becoming crucial.

Although the radio source is undoubtedly variable on short timescales
(3C\,84 in NGC\,1275 has varied considerably over the past 50yr), the
heating/cooling balance remains fairly tight on timescales longer than
$10^8\yr$ and must be long lived (several Gyr) in most objects.
This means it lasts over tens to hundreds of bubble cycles. Quite how
this is achieved is unclear, especially when the large range of
distances involved in the feedback cycle is considered.  For a billion
solar mass black hole the event horizon (near which most of the energy
is released) is about $10^{-4}\pc$, the Bondi accretion radius (which
may control the black hole accretion rate) may be 30~pc, the gas below
1~keV is mostly within a few kpc and the cooling radius (where the
radiative cooling time of the gas is say 7~Gyr) may be
100~kpc. Angular momentum should spoil any simple feedback loop,
introducing an uncertain and possibly long delay. Perhaps this can be
overcome if the gas is really viscous, transporting the angular
momentum outward?

In summary, the mechanisms by which energy is fed back into the
intracluster medium from a central black hole are observed. A close
heating/cooling balance has been set up which is not fully understood.
There remains the possibility, however, that there is continuous mild
overheating, where the balance is tipped slightly towards heating.


\begin{theacknowledgments}
 Thanks to Helen Russell, Roderick Johnstone and our many collaborators. 
\end{theacknowledgments}



\bibliographystyle{aipproc}   

\bibliography{refs}

\IfFileExists{\jobname.bbl}{}
 {\typeout{}
  \typeout{******************************************}
  \typeout{** Please run "bibtex \jobname" to optain}
  \typeout{** the bibliography and then re-run LaTeX}
  \typeout{** twice to fix the references!}
  \typeout{******************************************}
  \typeout{}
 }

\end{document}